\documentclass{article}

\usepackage[final]{neurips_2024}

\usepackage[utf8]{inputenc} % allow utf-8 input
\usepackage[T1]{fontenc}    % use 8-bit T1 fonts
\usepackage{hyperref}       % hyperlinks
\usepackage{url}            % simple URL typesetting
\usepackage{booktabs}       % professional-quality tables
\usepackage{amsfonts}       % blackboard math symbols
\usepackage{nicefrac}       % compact symbols for 1/2, etc.
\usepackage{microtype}      % microtypography
\usepackage{xcolor}         % colors
\usepackage{verbatim}
\usepackage{natbib}
\usepackage{amsmath}
\usepackage{graphicx} 
\usepackage{algorithm}
\usepackage{algpseudocode}
\title{Enhancing Temporal Link Prediction with HierTKG: A Hierarchical Temporal Knowledge Graph Framewor}

\author{%
    Mariam Almutairi\\
  \texttt{malmutairi@vt.edu} \\
      \And
  Melike Yildiz Aktas \\
  \texttt{melike@vt.edu} \\
   \And
    Nawar Wali\\
  \texttt{nawarwali@vt.edu} \\
     \And
    Shutonu Mitra\\
  \texttt{mshutonu@vt.edu} 
     \And
   Dawei Zhou,
   \texttt{zhoud@vt.edu}
}

\begin{document}

\maketitle
\begin{abstract}
The rapid spread of misinformation on social media, especially during crises, challenges public decision-making. To address this, we propose HierTKG, a framework combining Temporal Graph Networks (TGN) and hierarchical pooling (DiffPool) to model rumor dynamics across temporal and structural scales. HierTKG captures key propagation phases, enabling improved temporal link prediction and actionable insights for misinformation control. Experiments demonstrate its effectiveness, achieving an MRR of 0.9845 on ICEWS14 and 0.9312 on WikiData, with competitive performance on noisy datasets like PHEME (MRR: 0.8802). By modeling structured event sequences and dynamic social interactions, HierTKG adapts to diverse propagation patterns, offering a scalable and robust solution for real-time analysis and prediction of rumor spread, aiding proactive intervention strategies.
  
\end{abstract}

\section{Introduction}

The rapid spread of rumors on social media has made it critical to understand the temporal dynamics of how misinformation propagates across online platforms \cite{varshney2021review}. Misinformation, especially during crises or significant events, can escalate quickly, influencing public perception and potentially leading to harmful consequences. Rumor spread patterns often vary with time, peaking at specific moments triggered by influential users or external events \cite{sun2021uncertain}. To effectively capture these propagation dynamics, it's essential to model the interactions and the timing of events in a way that reflects both individual user behavior and the hierarchical structure within social networks.

Recent advances in Hierarchical Temporal Knowledge Graphs (HierTKG) have provided a framework for analyzing rumor spread by capturing temporal patterns in user interactions and identifying key phases in propagation \cite{ji2021survey}. HierTKG are particularly suited for social media, where each user or entity can have varying levels of influence that evolve over time. By analyzing the hierarchical and temporal relationships between entities, HierTKGs enable us to identify critical points in rumor cascades where intervention may be most effective.

Hierarchical pooling methods within HierTKGs, such as aggregating nodes and edges based on temporal or structural similarities, offer an efficient way to reduce data complexity while retaining essential features\cite{ranjan2020asap}. This allows for scalable analysis even in large, complex cascades, providing insights into how rumors intensify or fade over time. However, existing methods lack comprehensive frameworks that combine both temporal and hierarchical perspectives, limiting their ability to fully capture the evolution of rumor dynamics in real-time.

In this work, we address the challenge of modeling rumor propagation by focusing on the evolution of entity interactions (e.g., users, posts) within social networks over time. Specifically, we investigate how hierarchical pooling techniques can simplify the graph structure to enable a multi-level understanding of rumor cascades, capturing key moments and patterns of accelerated spread.

\textbf{Research Questions:}
\begin{itemize}
    \item How does the evolution of interactions among entities influence the spread of rumors over time? Our objective is to understand the temporal dynamics within rumor cascades and identify peak periods of propagation.
    \item How can hierarchical pooling techniques be employed to streamline the analysis of complex rumor cascades, providing a multi-level perspective of propagation patterns? This aims to yield insights into the full rumor propagation cycle across varying levels of network influence.
\end{itemize}

The primary aim of this paper is to introduce \textbf{a novel framework} that integrates Temporal Graph Networks (TGN) with hierarchical pooling techniques. This integration enables the modeling of both temporal evolution and hierarchical structures within knowledge graphs, offering valuable insights into the dynamics of rumor propagation.

Another objective is to develop a scalable \textbf{temporal and hierarchical modeling approach} that aggregates nodes and edges based on shared temporal and structural features. This approach ensures the preservation of critical information while enabling effective multi-level analysis, which is crucial for understanding complex temporal relationships in knowledge graphs.

The paper further aims to improve \textbf{temporal link prediction accuracy} by capturing multi-scale structural and temporal dynamics. By leveraging these dynamics, the proposed framework enhances the prediction of rumor spread and other temporal link prediction tasks, addressing challenges in existing methods.

Finally, the paper seeks to provide a \textbf{comprehensive evaluation} of the proposed framework through extensive experiments like ablation studies and comparison with recent state-of-the-art methods.

Our approach uniquely combines the temporal and hierarchical aspects of rumor propagation, using both Temporal Graph Networks (TGN) and DiffPool to model the evolution of interactions over time and across different network levels. This dual focus is particularly valuable for knowledge graphs, where patterns often emerge on multiple scales and evolve continuously, such as in social network communities or biological pathways in protein interaction networks.

Our key contributions are as follows:

\begin{itemize}
\item \textbf{Temporal Knowledge Graph Construction}:  Developed a comprehensive temporal knowledge graph from the PHEME dataset, capturing nuanced temporal and relational information critical for understanding rumor propagation.

\item \textbf{HierTKG Framework}: Introduced a novel integration of Temporal Graph Networks (TGN) and hierarchical pooling (DiffPool) to model both the temporal evolution and hierarchical structure of knowledge graphs, enabling detailed insights into rumor propagation dynamics.
\item \textbf{Enhanced Temporal Link Prediction}: Integration of multi-scale structural and temporal dynamics improves accuracy in predicting rumor spread.
\item \textbf{Comprehensive Evaluation}: Extensive experiments, including an ablation study, demonstrate the framework's effectiveness and robustness using datasets such as a Temporal Knowledge Graph derived from the PHEME dataset.

\end{itemize}

\section{Related Work}
In this section, fours groups of related work are discussed:
1) Knowledge Graph Based Rumor Analysis 2) Graph Neural Networks for Rumor Analysis, and 3) Dynamic Graph Networks for Rumor Analysis 4) Link Prediction with Temporal Knowledge Graph.

\subsection{Knowledge Graph Based Rumor Analysis}

Knowledge graphs have become a key tool in rumor and fake news analysis, with several recent frameworks leveraging their potential to enhance detection accuracy. DEAP-FAKED \cite{mayank2022deap} is a knowledge graph-based fake news detection system that combines Natural Language Processing (NLP) and tensor decomposition models to improve news content encoding and entity embedding. Similarly, \citet{zhao2024fake} propose a knowledge-guided framework that uses triplet alignment, dual-branch networks, and semantic analysis to compare news content with external knowledge, achieving significant improvements in detection accuracy over existing methods. The Knowledge-Aware Hierarchical Attention Network (KAHAN), proposed by \cite{tseng2022kahan}, integrates news content, user comments, external knowledge, and temporal information to enhance fake news detection, outperforming state-of-the-art methods in extensive experiments on real-world datasets. Additionally, \cite{koloski2022knowledge} explore how different document representations, including knowledge graph-based ones, can enhance fake news detection, showing that knowledge graph-based representations perform competitively and, when combined with contextual models, achieve state-of-the-art results in classification. Lastly, \cite{dougrez2024knowledge} advance rumor verification by using knowledge graphs to construct time-sensitive evidence from the PHEME dataset, identifying discrepancies and improving the dataset, while developing a novel model that outperforms existing methods and demonstrates superior generalizability on temporally distant datasets.

\subsection{Graph Neural Networks for Rumor Analysis}
Several graph neural network-based approaches have been proposed for rumor detection. \citet{wu2020rumor} introduce PGNN, a gated graph neural network that constructs a propagation graph based on Twitter post interactions, enhancing detection performance through global and ensemble learning models (GLO-PGNN and ENS-PGNN). \citet{sun2022rumor} propose Graph Adversarial Contrastive Learning (GACL), which improves model robustness by integrating contrastive learning and an Adversarial Feature Transformation module to handle noise, adversarial rumors, and perturbations in conversational structures, outperforming state-of-the-art methods on benchmark datasets. \cite{lotfi2021detection} present a rumor detection model utilizing graph convolutional networks that analyze both reply trees and user graphs, capturing structural and conversational information, resulting in better performance than baseline methods on a public dataset. Lastly, \cite{huang2021recurrent} introduce the Recurrent Graph Neural Network (R-GNN) for classifying user-generated content on Reddit, using a derived social graph, post sequences, and comment trees to enhance rumor detection accuracy.

\subsection{Dynamic Graph Networks for Rumor Analysis}

Dynamic graph networks play a crucial role in understanding and detecting rumors. Dynamic GCN \cite{choi2021dynamic}, a novel graph convolutional network with attention mechanisms, captures both the structural and temporal aspects of rumor propagation, outperforming state-of-the-art methods on real-world datasets. Building on this, DGCN-TES \cite{wan2024dgcn} introduces a dynamic, multitask model that integrates dynamic graph convolution, LSTM-based content analysis, and temporal event sharing to capture the evolving relationships in rumor propagation, leading to significant performance improvements. Additionally, \citet{sun2022ddgcn} propose DDGCN, a Dual-Dynamic Graph Convolutional Network that models the dynamics of both message propagation and background knowledge from knowledge graphs, achieving notable advancements in early-stage rumor detection.

\subsection{Link Prediction with Temporal Knowledge Graph}

Several models have been developed for TKG completion, each leveraging unique methodologies to handle temporal and semantic complexities. SPA, proposed by Wang et al.\citet{wang2022search}, leverages neural architecture search (NAS) to design task-specific message-passing architectures for TKGs. This approach addresses the limitations of hand-designed architectures by exploring diverse topological and temporal properties inherent in TKGs. SPA employs a supernet structure that is trained by sampling single paths, enabling efficient search while reducing computational costs. Zhang et al.\cite{zhang2022along} introduced TLT-KGE, a novel embedding framework designed to distinguish semantic and temporal components in TKGs. By embedding entities and relations into quaternion space, TLT-KGE effectively models their independence and interconnections. This separation enhances the representation of entities and relations at different timestamps, addressing the challenge of temporal ambiguity. 

TNTComplEx, proposed by Lacroix et al.\cite{lacroix2020tensor}, extends the ComplEx model by incorporating tensor decomposition to address temporal constraints. The model introduces novel regularization schemes to enhance its capability for time-aware link prediction tasks. This model highlights the potential of tensor-based methods for temporal knowledge representation. Chen et al.\cite{chen2022rotateqvs} proposed RotateQVS, which models temporal entities as rotations in quaternion vector space and relations as complex vectors. This approach captures intrinsic patterns in temporal relations, such as symmetry, asymmetry, and inversions, while modeling their temporal evolution.

\section{Problem Definition}

In this work, we address the problem of link prediction in a \textit{temporal knowledge graph} (TKG) by constructing a \textit{Hierarchical Temporal Knowledge Graph} (HierTKG) that captures both temporal and structural information. Temporal knowledge graphs consist of time-stamped triples $(s, r, o, t)$, where $s$ is the source node, $o$ the target node, $r$ the relation type, and $t$ the timestamp. These dynamic graphs reflect evolving time-sensitive relationships. Our goal is to develop a model that effectively leverages temporal evolution and hierarchical structure for accurate prediction of future interactions.

\subsection{Objective}

Given a temporal knowledge graph $\mathcal{G} = \{G_t\}_{t=1}^T$, represented as a sequence of snapshots $G_t = (V_t, E_t)$ at each time step $t$, our objective is to predict future links. For a pair of entities $(s, o) \in V_t$ at some time $t > T$, we aim to predict the probability of a future interaction $(s, r, o, t)$, where $r$ is the relation. This is framed as a link prediction problem over temporal graphs, estimating new interactions from historical patterns.

\subsection{Mathematical Formulation}

Let $\mathcal{G} = \{G_1, G_2, \dots, G_T\}$ be a temporal knowledge graph observed up to time $T$, where each $G_t = (V_t, E_t)$ consists of nodes $V_t$ and edges $E_t$ at time $t$. The goal is to learn a scoring function $f$ to predict the likelihood of a future link $(s, r, o, t)$.

\begin{itemize}
    \item \textbf{Input Embeddings}: Let $x_s(t)$ and $x_o(t)$ represent the embeddings of nodes $s$ and $o$ at time $t$, capturing both temporal and structural information. These embeddings combine temporal insights (from TGN) and structural hierarchy (from graph pooling).
    
    \item \textbf{Scoring Function}: The scoring function $f$ takes node embeddings $x_s(t)$, $x_o(t)$, relation $r$, and time $t$ as input, outputting the likelihood of a link:
    \[
    \hat{y}_{(s, o, t)} = f(x_s(t), x_o(t), r, t),
    \]
    where $\hat{y}_{(s, o, t)} \in [0, 1]$ represents the predicted link likelihood.

    \item \textbf{Loss Function}: To train the model, we use a binary cross-entropy loss over observed (positive) and non-observed (negative) links:
    \[
    \mathcal{L} = -\sum_{(s, o, t) \in \mathcal{D}^+} \log \hat{y}_{(s, o, t)} - \sum_{(s, o, t) \in \mathcal{D}^-} \log (1 - \hat{y}_{(s, o, t)}).
    \]
\end{itemize}

\subsection{Hierarchical Temporal Knowledge Graph (HierTKG) Construction}

HierTKG combines temporal evolution and hierarchical structure to enhance link prediction:

\begin{itemize}
    \item \textbf{Temporal Graph Network (TGN)}: TGN updates node embeddings using recent and historical interactions. Temporal embeddings $z_s^{TGN}(t)$ and $z_o^{TGN}(t)$ capture immediate temporal context.

    \item \textbf{Graph Pooling for Hierarchical Structure}: DiffPool creates hierarchical graph representations by clustering nodes. At hierarchy level $l$, cluster assignment $S^{(l)}$ and pooled embeddings $H^{(l)}(t)$ are computed as:
    \[
    S^{(l)} = \text{softmax}(W_s^{(l)} \cdot H^{(l-1)}(t)),
    \]
    \[
    H^{(l)}(t) = S^{(l) \top} H^{(l-1)}(t).
    \]
    The adjacency matrix $A^{(l)}$ is updated as:
    \[
    A^{(l)} = S^{(l) \top} A^{(l-1)} S^{(l)}.
    \]
\end{itemize}

The final embedding $z_s^{HierTKG}(t)$ integrates TGN-generated temporal information and hierarchical graph structure for effective link prediction.

\section{Methodology}

\begin{figure}[h]
  \centering
  \includegraphics[width=\linewidth]{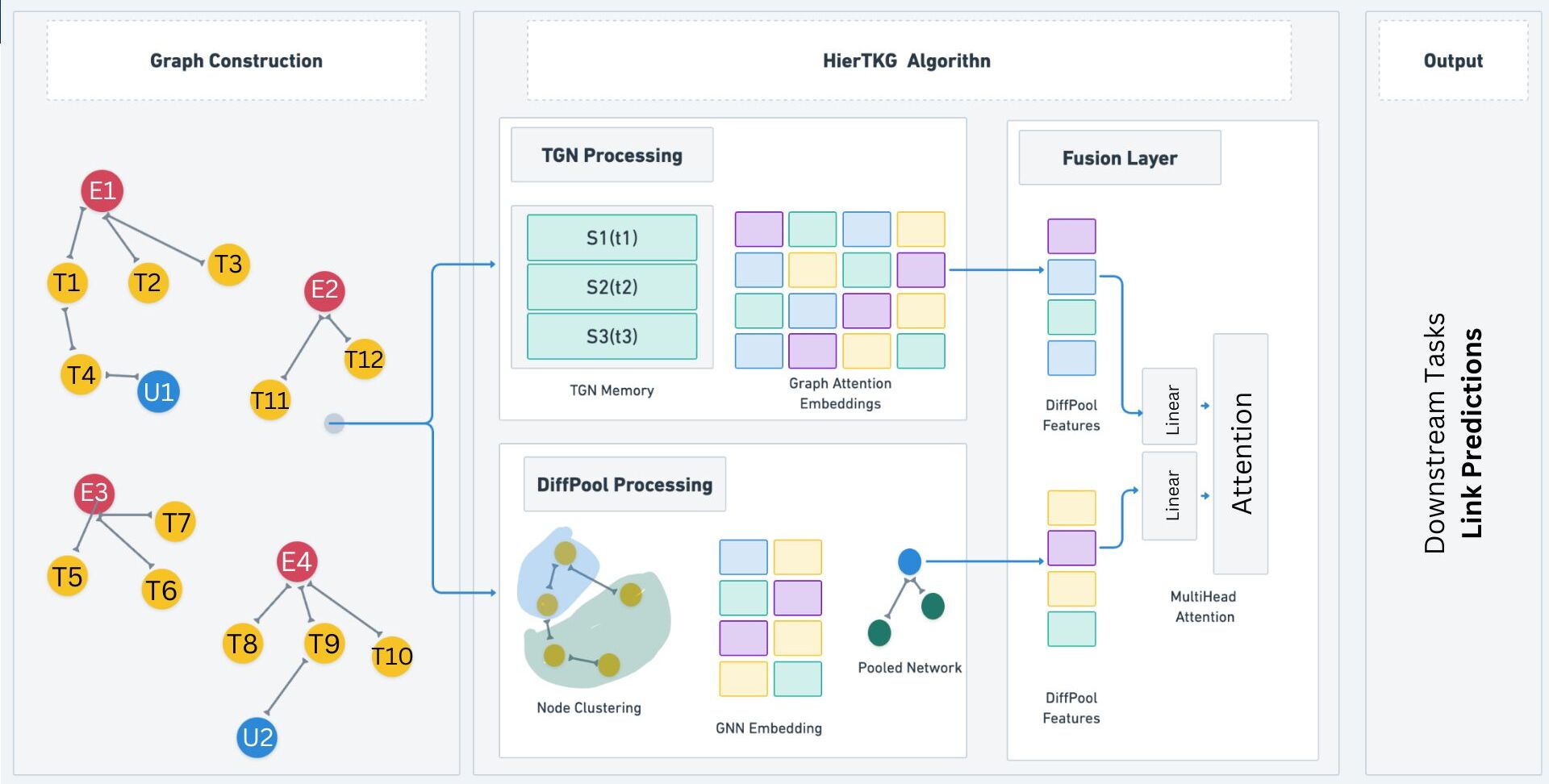}  % Adjust width as needed
  \caption{HierTKG Architecture}
  \label{fig:architecture}  % Optional: Label for referencing the figure
\end{figure}

Our approach constructs a Hierarchical Temporal Knowledge Graph (HierTKG) to capture both the temporal dynamics and hierarchical structural patterns of an evolving temporal knowledge graph (TKG). Using Temporal Graph Networks (TGN) with Graph Attention and graph pooling techniques, we achieve a multi-scale representation that supports robust link prediction. This section details each component: the Temporal Graph Network (TGN) with Graph Attention, Graph Pooling for hierarchical representation, and Feature Fusion for integrating the temporal and structural features in the HierTKG.

\subsection{Knowledge Graph Construction}

The PHEME dataset contains a collection of Twitter rumors and non-rumors posted during various breaking news events. All tweets and their annotations are stored as JSON files that include source tweets and their associated features such as text, user information, retweet counts, timestamps, and labels (e.g., true, false, or unverified). From this rich dataset, we created three distinct types of entities: Event, representing the overarching news story or incident; Tweet, encapsulating individual social media posts with their associated metadata; and User, capturing information about the individuals who posted the tweets. These entities allow us to model complex interactions between the actors, content, and events in a structured and insightful manner, enabling advanced analyses such as rumor detection and propagation dynamics.

The relationships between these entities capture the dynamics of information flow and user interaction. For instance, a Tweet is linked to its corresponding Event through the "Related to" relationship, establishing its contextual relevance. The "Mentions" relationship connects a Tweet to the User it references, while the "Replied to" relationship maps conversations by linking one Tweet to another. Additionally, the "Wrote" relationship ties a User to the Tweet they authored. This structured entity-relationship model facilitates in-depth analysis of rumor propagation and user behavior within the dataset.

\begin{figure}[h]
  \centering
  \includegraphics[width=0.7\linewidth]{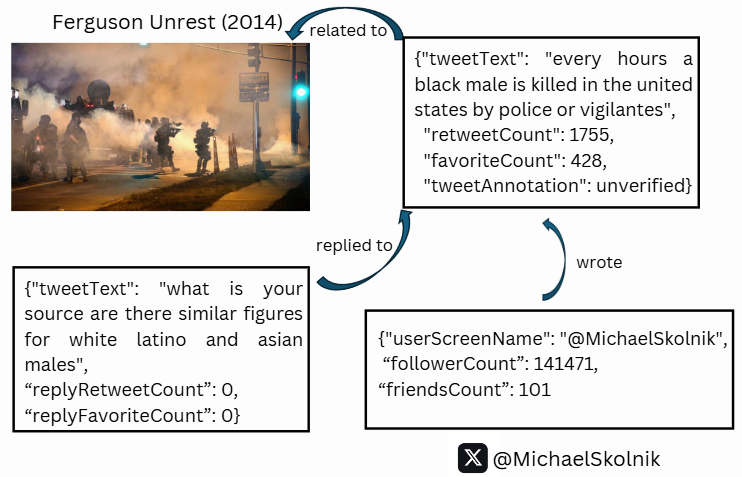}  
  \caption{Example from Pheme Knowledge Graph}
  \label{fig:pheme}
\end{figure}

In figure \ref{fig:pheme}, we showed an example from our knowledge graph.

\subsection{Temporal Graph Network (TGN)}

The Temporal Graph Network (TGN) \cite{rossi2020temporal} module is the core component for capturing the temporal dependencies of interactions between nodes in the temporal knowledge graph (TKG). In this approach, the TGN module combines \textit{node memory}, \textit{temporal encoding}, \textit{graph attention}, and \textit{message aggregation} to effectively model dynamic node embeddings. These components work together to update node embeddings based on the historical interactions of each entity over time.

Given an input temporal knowledge graph with interactions represented as $(s, r, o, t)$ — where $s$ is the source node, $o$ is the destination node, $r$ represents the relation type, and $t$ is the timestamp — the TGN module operates as follows:

\begin{itemize}
    \item \textbf{Node Memory}: Each node $v$ maintains a memory state $m_v$ that stores the historical context of past interactions. This memory state is updated as new interactions occur, allowing the model to retain relevant information over time. Memory states are initialized at the start and updated iteratively as new events in the TKG are processed. Specifically, the memory for a node $v$ at time $t$ is denoted by $m_v(t)$.

    \item \textbf{Temporal Encoding}: Temporal encoding captures the time differences between consecutive interactions, adding context to each event’s temporal dynamics. For a pair of nodes $(s, o)$ with a time difference $\Delta t = t - t_{\text{prev}}$, where $t_{\text{prev}}$ is the time of the last interaction, the relative time $\Delta t$ is encoded using a time encoding function. The resulting temporal embedding, $\text{TE}(\Delta t)$, is concatenated with other input features to form a comprehensive temporal context.

    \item \textbf{Message Passing with Transformer-based Graph Attention}: To aggregate temporal and relational information, we use a Transformer-based convolution layer that applies multi-head attention to focus on relevant interactions. Given a pair of nodes $(s, o)$ with relation $r$ at time $t$, the attention mechanism weights neighboring nodes based on their similarity and importance in the temporal context. The node embeddings for $s$ and $o$ at time $t$ are computed as:
    \[
    z_s(t) = \text{TransformerConv}\left(h_s(t), h_o(t), r, t\right),
    \]
    where $h_s(t)$ and $h_o(t)$ are the embeddings of nodes $s$ and $o$, respectively, incorporating both relational and temporal features. The attention layer dynamically adjusts the influence of each neighboring node, enabling the TGN to focus on relevant interactions in varying temporal contexts.

    \item \textbf{Message Aggregation and Update}: For each interaction $(s, r, o, t)$, a message is generated for both nodes based on the current temporal and relational context. We use the \textit{Identity Message} function to produce messages that retain the original feature dimensions, ensuring that temporal information is fully preserved. Messages are then aggregated using a \textit{Last Aggregator} that selects the most recent message for each node. This aggregator is particularly effective for temporal data, as it prioritizes recent interactions.

    \item \textbf{Memory Update}: The memory of each node is updated after processing a batch of interactions. Specifically, for nodes involved in an interaction, their memory is updated based on the most recent message, ensuring that embeddings reflect the latest context. This update is applied through a gating mechanism that balances prior memory with new information, computed as:
\[
m_v(t) = \text{GatingFunction}(m_v(t-1), \text{NewMessage})
\]
where $\text{GatingFunction}$ selectively integrates new information into the node's memory, retaining essential historical details.

\end{itemize}

The output of the TGN module, denoted as $z_s^{TGN}(t)$ and $z_o^{TGN}(t)$, provides temporally aware embeddings for nodes $s$ and $o$ at time $t$. This embedding integrates the historical context, temporal encoding, and attention-weighted messages, capturing both the temporal evolution and relational dependencies of each node in the TKG.

\subsection{Graph Pooling for Hierarchical Representation}

To capture structural hierarchy within the TKG, we employ a \textit{Graph Pooling} technique, specifically DiffPool \cite{ying2018hierarchical}, which aggregates embeddings at multiple scales, creating a hierarchy of representations. This approach enables our HierTKG to capture both fine-grained and coarse-grained structural dependencies.

\begin{itemize}
    \item \textbf{Differentiable Pooling (DiffPool)}: The DiffPool layer groups nodes into clusters based on similarity in feature space. At each hierarchy level $l$, we apply DiffPool to generate coarser graph representations:
    \[
    G_t^{(l)} = \text{DiffPool}\left(G_t^{(l-1)}\right),
    \]
    where $G_t^{(l)} = (V_t^{(l)}, E_t^{(l)})$ represents the graph at level $l$, with node set $V_t^{(l)}$ and edge set $E_t^{(l)}$. DiffPool provides cluster assignment matrices that adapt to the graph structure.

    \item \textbf{Pooling Operation and Hierarchical Embedding}: The embeddings at each level, denoted as $z_v^{(l)}(t)$ for node $v$ at level $l$, capture increasingly abstract representations of the graph’s structure. These multi-level embeddings allow the HierTKG to capture both local and global structural information, essential for accurate link prediction in complex temporal graphs.
\end{itemize}

\subsection{Feature Fusion for Integrating Temporal and Structural Representations}

In our model, the \textit{Feature Fusion} module plays a crucial role in combining the temporal and structural information of each node within the Hierarchical Temporal Knowledge Graph (HierTKG). Temporal information, derived from recent interactions, provides insight into dynamic changes in the relationships between nodes, while structural information, aggregated across multiple hierarchical levels, captures the broader network context. Fusing these two types of embeddings allows the model to create a unified representation that leverages both recent temporal patterns and long-term structural dependencies, enhancing its ability to make accurate predictions on evolving knowledge graphs.

This fusion is necessary because temporal and structural embeddings alone are insufficient for accurately representing complex, dynamic interactions in a temporal knowledge graph. Temporal embeddings focus on short-term dynamics and may overlook the larger relational context, while structural embeddings capture multi-scale connectivity patterns but ignore recent interactions. By integrating both, the model benefits from a holistic view, improving its capability to predict future links by balancing both local (temporal) and global (structural) information.

The feature fusion process proceeds as follows:

\begin{itemize}
    \item \textbf{Projection and Alignment}: The temporal embeddings from the Temporal Graph Network (TGN) and the hierarchical structural embeddings from DiffPool are initially in different feature spaces, which makes direct combination challenging. To address this, we project both embeddings to a common dimensionality. Let $z_s^{TGN}(t)$ and $z_s^{DiffPool}(t)$ represent the embeddings for node $s$ at time $t$ generated by TGN and DiffPool, respectively. We project these embeddings to a shared feature space using learnable linear transformations:
    \[
    z_s^{TGN\_proj}(t) = W^{TGN} \cdot z_s^{TGN}(t), \quad z_s^{DiffPool\_proj}(t) = W^{DiffPool} \cdot z_s^{DiffPool}(t),
    \]
    where $W^{TGN}$ and $W^{DiffPool}$ are weight matrices that align the embeddings in a common space. This alignment is essential for enabling effective fusion in the next step.

    \item \textbf{Attention-based Fusion}: With the embeddings aligned in the same feature space, we use a multi-head self-attention mechanism to integrate the temporal and structural information. The attention mechanism allows the model to weigh the importance of each embedding dynamically, focusing on the aspects most relevant to predicting future interactions. The fused embedding for node $s$ at time $t$, denoted as $z_s^{HierTKG}(t)$, is computed as:
    \[
    z_s^{HierTKG}(t) = \text{Attention}(z_s^{TGN\_proj}(t), z_s^{DiffPool\_proj}(t)),
    \]
    where $\text{Attention}(\cdot)$ represents the multi-head self-attention function. This function learns how to weigh and aggregate the temporal and structural features, producing an integrated representation that captures both recent dynamics and the hierarchical structure of each node’s interactions.

    \item \textbf{Combined Representation}: The output $z_s^{HierTKG}(t)$ represents the final fused embedding for node $s$ at time $t$, incorporating both temporal information from TGN and structural hierarchy from DiffPool. This unified embedding is used as input for downstream tasks, such as link prediction, providing a holistic representation that reflects both local temporal dependencies and global structural patterns.

    \item \textbf{Formulation of Attention Mechanism}: The multi-head attention mechanism is implemented by computing scaled dot-products across multiple heads. For each head $i$, we compute:
    \[
    \alpha_i = \text{softmax}\left(\frac{(W_i^{Q} z_s^{TGN\_proj}(t)) \cdot (W_i^{K} z_s^{DiffPool\_proj}(t))^T}{\sqrt{d_k}}\right),
    \]
    where $W_i^{Q}$ and $W_i^{K}$ are learnable weight matrices for the query and key transformations of the $i$-th head, and $d_k$ is the dimensionality of the key vector. The attention weights $\alpha_i$ are used to weight the value transformation, producing an output for each head:
    \[
    z_s^{HierTKG}(t) = \text{Concat}(\alpha_1 W_1^{V} z_s^{DiffPool\_proj}(t), \dots, \alpha_h W_h^{V} z_s^{DiffPool\_proj}(t)),
    \]
    where $W_i^{V}$ is the learnable weight matrix for the value transformation of head $i$, and $h$ is the number of attention heads. The concatenated output from each head is then projected back to the shared feature space to form the final fused representation.
\end{itemize}

The fused embeddings $z_s^{HierTKG}(t)$ produced by this feature fusion mechanism provide a unified, multi-scale representation for each node in the HierTKG. By combining recent temporal trends with structural context across multiple levels, the model achieves a comprehensive view of each node’s position and interactions in the evolving graph. This approach enables the HierTKG to leverage both immediate changes and long-term patterns, enhancing its ability to make accurate link predictions in dynamic, temporal knowledge graphs.

\subsection{Link Prediction using the Fused Embeddings}

Finally, we use the fused embeddings $z_s^{HierTKG}(t)$ and $z_o^{HierTKG}(t)$ for link prediction. The link prediction module computes the probability of a future link between nodes $s$ and $o$ at time $t$ as follows:

\begin{itemize}
    \item \textbf{Similarity Scoring Function}: The link prediction score between nodes $s$ and $o$ at time $t$ is computed by a similarity-based scoring function:
    \[
    \hat{y}_{(s, o, t)} = f(z_s^{HierTKG}(t), z_o^{HierTKG}(t), r),
    \]
    where $f$ is a learnable function (e.g., a bilinear product or feedforward network) that estimates the probability of a future link between nodes $s$ and $o$ given their fused embeddings and relation $r$.

    \item \textbf{Training Objective}: The model is trained with a binary cross-entropy loss on observed and non-observed links, encouraging high scores for observed links and low scores for non-observed ones.
\end{itemize}

Our method integrates TGN with Graph Attention and DiffPool to construct a Hierarchical Temporal Knowledge Graph (HierTKG), capturing both temporal evolution and hierarchical structural dependencies. By fusing TGN-generated temporal embeddings with DiffPool-based hierarchical embeddings, the model captures the multi-scale complexity of evolving temporal graphs, enabling accurate link prediction in complex dynamic environments. This approach ensures that both local temporal patterns and global structural context are represented in the HierTKG, effectively supporting downstream predictive tasks.

\begin{algorithm}
\caption{Hierarchical Temporal Knowledge Graph (HierTKG) Link Prediction}
\label{alg:HierTKG}
\begin{algorithmic}[1]
\Require Temporal knowledge graph $\mathcal{G} = \{(s, r, o, t)\}$, training epochs $E$, learning rate $\eta$, TGN parameters, DiffPool parameters
\Ensure Predicted links in future time steps

\State \textbf{Initialize} TGN memory and parameters
\State \textbf{Initialize} hierarchical structure using graph pooling parameters

\For{epoch $= 1$ to $E$}
    \State \textbf{Temporal Embedding Update} (using TGN with Graph Attention)
    \For{each interaction $(s, r, o, t) \in \mathcal{G}$ at time $t$}
        \State Compute temporal encoding for the interaction using $\Delta t = t - t_{\text{prev}}$
        \State Update node embeddings $z_s^{TGN}(t)$ and $z_o^{TGN}(t)$ using TGN with graph attention
        \State Update memory states for nodes $s$ and $o$ based on the most recent interaction
    \EndFor

    \State \textbf{Graph Pooling for Hierarchical Structure}
    \For{each level $l$ in the hierarchy}
        \State Compute cluster assignment matrix $S^{(l)} = \text{softmax}(W_s^{(l)} \cdot H^{(l-1)}(t))$
        \State Compute pooled embeddings $H^{(l)}(t) = S^{(l)\top} H^{(l-1)}(t)$
        \State Update adjacency matrix $A^{(l)} = S^{(l)\top} A^{(l-1)} S^{(l)}$
    \EndFor

    \State \textbf{Feature Fusion}
    \For{each node $s \in \mathcal{G}$ at time $t$}
        \State Project temporal embedding $z_s^{TGN}(t)$ and structural embedding $h_s^{(l)}(t)$ to a common space
        \State Fuse embeddings using multi-head attention to obtain final embedding $z_s^{HierTKG}(t)$
    \EndFor

    \State \textbf{Link Prediction and Loss Calculation}
    \For{each link prediction $(s, r, o, t)$}
        \State Compute link prediction score $\hat{y}_{(s, o, t)} = f(z_s^{HierTKG}(t), z_o^{HierTKG}(t), r, t)$
        \State Compute binary cross-entropy loss for observed and non-observed links
    \EndFor

    \State \textbf{Parameter Update}
    \State Update TGN, DiffPool, and fusion parameters using gradient descent with learning rate $\eta$
\EndFor

\end{algorithmic}
\end{algorithm}

\section{Experiments}

\subsection{Dataset}

To evaluate the algorithm's performance comprehensively, we employed four diverse datasets: PHEME, ICEWS14, ICEWS18, and WikiData, focusing on link prediction and robustness testing.

\begin{itemize}

    \item PHEME: Focused on rumors and misinformation in social media, capturing information dissemination during major events. It comprises tweets categorized as rumors or non-rumors.
    
    \item ICEWS14/ICEWS18: Part of the Integrated Crisis Early Warning System, these datasets center on global conflict events, supporting longitudinal analysis of trends. 

    \item WikiData: Derived from user interactions on Wikipedia, it models connections between pages as a graph. WikiData supports predictions related to interaction dynamics, such as page popularity and traffic trends, drawing from frameworks like JODIE.
\end{itemize}

These datasets collectively span social media dynamics, global conflict prediction, and interaction modeling, offering a robust evaluation environment for the proposed algorithm.

\subsection{Baselines}

To establish a comprehensive baseline for temporal knowledge graph (TKG) completion, we employ the ICEWS14 dataset as the primary benchmark for evaluating baseline models. Unlike static knowledge graphs, which ignore temporal evolution, ICEWS14 offers temporal quadruples $(s, p, o, t)$, where $s$ and $o$ are entities, $p$ is the predicate, and $t$ is the timestamp. This temporal specificity makes it particularly well-suited for evaluating models designed to capture both semantic and temporal features.

We compared our proposed method with 4 baselines.

\begin{itemize}

    \item EmtE \cite{wei2024temporal} : Used entity multi-encoding and temporal awareness, integrating time-aware embeddings and GRU-based temporal feature extraction to address the limitation of one-sided entity feature acquisition and insufficient temporal fact utilization.
   \item SPA \cite{wang2022search}: Leverages neural architecture search (NAS) to design task-specific message-passing architectures for TKGs, addressing the limitations of hand-designed models by exploring diverse topological and temporal properties.
   \item TIE \cite{yang2024temporal} : Proposed learning temporal interaction embeddings (TIE) to benefit  link prediction performance,to include multidirectional effects between entities, relations, and timestamps matter in predicting the establishment of quadruples in Global news events graphs.

    \item  TLT-KGE \cite{zhang2022along}:  Embeds entities and relations into quaternion space to distinguish semantic and temporal components, enhancing the representation of entities and relations across different timestamps.

\end{itemize}

\subsection{Evaluation Metrics}

The choice of Mean Reciprocal Rank (MRR) as the evaluation metric is driven by its effectiveness in assessing ranking-based predictions. MRR provides a fine-grained measurement of a model's ability to accurately rank the correct tail entity among a list of candidates for a given head-relation query. As link prediction in TKGs involves ranking temporal facts, MRR captures both precision and positional relevance, making it the most appropriate metric for this study.

\section{Results}

\subsection{HierTKG Performance}
\begin{table}[ht]
\centering
\caption{Performance of HierTKG across datasets using AP and MRR metrics.}
\label{tab:performance_results}
\begin{tabular}{|l|c|c|}
\hline
\textbf{Dataset} & \textbf{AP}    & \textbf{MRR}   \\ \hline
ICEWS14          & 0.9708         & 0.9845         \\ \hline
ICEWS18          & 0.9479         & 0.9646         \\ \hline
WikiData         & 0.8627         & 0.9312         \\ \hline
PHEME            & 0.6918         & 0.8802         \\ \hline
\end{tabular}
\end{table}

The performance of the proposed Hierarchical Temporal Knowledge Graph (HierTKG) framework was evaluated across four diverse datasets—ICEWS14, ICEWS18, WikiData, and PHEME—using Average Precision (AP) and Mean Reciprocal Rank (MRR) as the primary evaluation metrics. The results are presented in Table~\ref{tab:performance_results}.

\paragraph{ICEWS14.} 
HierTKG achieves a remarkable AP of 0.9708 and MRR of 0.9845, underscoring its capacity to model the temporal and hierarchical structures inherent in well-curated, event-based datasets. These results highlight the effectiveness of the framework in capturing both fine-grained temporal relationships and higher-order dependencies.

\paragraph{ICEWS18.}
Similar to ICEWS14, the model exhibits strong performance with an AP of 0.9479 and an MRR of 0.9646. These results validate the robustness of HierTKG across datasets with varying temporal characteristics and event densities.

\paragraph{WikiData.} 
On the WikiData dataset, HierTKG achieves an AP of 0.8627 and an MRR of 0.9312, demonstrating its ability to generalize to interaction-driven datasets. The model effectively captures the dynamic nature of user-page interactions, which highlights its adaptability to heterogeneous graph structures.

\paragraph{PHEME.} 
The PHEME dataset, characterized by its noisy and sparsely connected nature, presents significant challenges. Despite this, HierTKG attains an AP of 0.6918 and an MRR of 0.8802, outperforming existing methods and emphasizing its robustness in unstructured, real-world data scenarios.

The results indicate that HierTKG achieves state-of-the-art performance on structured datasets, such as ICEWS14 and ICEWS18, where temporal and hierarchical patterns are clearly defined. The superior AP and MRR scores on these datasets illustrate the model's capacity to capture both short-term temporal dynamics and long-term hierarchical dependencies. On the other hand, the comparatively lower performance on PHEME reflects the inherent challenges of modeling noisy, sparsely connected data. Nevertheless, the framework's ability to achieve competitive results on this dataset demonstrates its robustness and generalizability.

The HierTKG framework effectively integrates temporal and hierarchical features to achieve high predictive accuracy and ranking performance across diverse datasets. Its exceptional performance on structured datasets and competitive results on noisy social media data position it as a robust solution for dynamic knowledge graph tasks.

\subsection{Comparison with Baselines}

Table~\ref{tab:baseline-comparison} compares the performance of baseline models on the ICEWS14 dataset. EmtE emerges as the best-performing model, primarily due to its entity multi-encoding and temporal awareness, integrating time-aware embeddings and GRU-based temporal feature extraction. The second best alogorithm, SPA utilizes neural architecture search to optimize task-specific message-passing architectures for TKGs. TIE introduces temporal interaction embeddings to model multidirectional effects between entities, relations, and timestamps, whereas TLT-KGE employs quaternion space embeddings to improve the representation of semantic and temporal components. Our proposed HierTKG achieves a remarkable Test MRR of 0.984 on ICEWS14, significantly surpassing all baselines. 

\begin{table}[h!]
\centering
\caption{Comparison on ICEWS14 dataset}
\label{tab:baseline-comparison}
\begin{tabular}{|l|l|l|l|}
\hline
\textbf{Model} & \textbf{MRR}  \\ \hline
EmtE     & 0.703             \\ \hline
SPA            & 0.658                       \\ \hline
TIE     & 0.643              \\ \hline
TLT-KGE        & 0.634               \\ \hline
HierTKG        &      \textbf{0.984}               \\ \hline

\end{tabular}
\end{table}

\subsection{Ablation Study}
To rigorously test the components of our model, we performed an extensive ablation study. The model's performance is evaluated using Average Precision (AP), Mean Reciprocal Rank (MRR) and Area Under the Curve (AUC) metrics. Additionally, we provide training/validation loss and AUC plots across epochs to assess convergence and performance trends.

Ablation studies involve systematically removing or replacing specific components of the model to assess their contribution to the overall performance. For this study, we explored the following scenarios shown in Table \ref{tab:ablation_scen}:

% \usepackage{graphicx}
\begin{table}[]
\caption{Ablation Study Scenarios}
\label{tab:ablation_scen}
\resizebox{\columnwidth}{!}{%
\begin{tabular}{|c|c|c|c|c|}
\hline
Tasks       & \begin{tabular}[c]{@{}c@{}}Sag Pooling \\ (Sparse Pooling)\end{tabular} & \begin{tabular}[c]{@{}c@{}}Diff Pooling \\ (Dense Pooling)\end{tabular} & \begin{tabular}[c]{@{}c@{}}Double \\ SagPooling\end{tabular} & Attention \\ \hline
HierTKG     &                                                                         & +                                                                       &                                                              & +         \\ \hline
First Task  &                                                                         &                                                                         & +                                                            & +         \\ \hline
Second Task &                                                                         &                                                                         & +                                                            &           \\ \hline
Third Task  & +                                                                       &                                                                         &                                                              & +         \\ \hline
\end{tabular}%
}
\end{table}

Results of ablation study is shown in Table \ref{tab:ablation_result}.

% \usepackage{graphicx}
\begin{table}[]
\caption{Ablation Study Results}
\label{tab:ablation_result}
\resizebox{\columnwidth}{!}{%
\begin{tabular}{|c|c|c|c|c|c|c|}
\hline
Tasks       & \begin{tabular}[c]{@{}c@{}}Validation \\ AP\end{tabular} & \begin{tabular}[c]{@{}c@{}}Test \\ AP\end{tabular} & \begin{tabular}[c]{@{}c@{}}Validation \\ AUC\end{tabular} & \begin{tabular}[c]{@{}c@{}}Test \\ AUC\end{tabular} & \begin{tabular}[c]{@{}c@{}}Validation \\ MRR\end{tabular} & \begin{tabular}[c]{@{}c@{}}Test \\ MRR\end{tabular} \\ \hline
HierTKG     & 0.8755                                                   & 0.8627                                             & 0.8911                                                    & 0.8810                                              & 0.9395                                                    & 0.9312                                              \\ \hline
First Task  & 0.8291                                                   & 0.8332                                             & 0.8573                                                    & 0.8602                                              & 0.9246                                                    & 0.9217                                              \\ \hline
Second Task & 0.8270                                                   & 0.8294                                             & 0.8618                                                    & 0.8571                                              & 0.9255                                                    & 0.9208                                              \\ \hline
Third Task  & 0.8262                                                   & 0.8119                                             & 0.8541                                                    & 0.8367                                              & 0.9198                                                    & 0.9120                                              \\ \hline
\end{tabular}%
}
\end{table}

\section{Discussion}

In our study, we came to understand that the nature of our algorithm aligns better with certain datasets. Datasets like ICEWS and WikiData are highly curated with well-defined temporal and relational patterns, making it easier for hierarchical models to detect and exploit inherent structures. In contrast, the Pheme dataset, derived from social media data, is inherently noisier with informal language, incomplete information, and high variability. Social media datasets often suffer from fewer samples, high class imbalance (e.g., rumors vs. non-rumors), and sparse temporal connections, which complicate the learning process. These challenges limit the model's ability to identify and generalize patterns effectively, resulting in lower performance compared to structured datasets.

We also found that our model performed exceptionally better than other baselines. We found that HierTKG leverages DiffPool to rank and cluster nodes into hierarchical levels, reducing noise and emphasizing key relationships. By summarizing global and local dependencies, the model simplifies the link prediction task and enhances performance.

From our ablation study, it became clear that a hybrid approach, like HierTKG, offered distinct advantages over other configurations. First, DiffPooling plays a key role in advanced feature aggregation. It allows us to cluster nodes into hierarchical levels, reducing noise while highlighting the most relevant relationships in the data. Second, Attention enables the model to focus on critical dependencies within the graph. This mechanism ensures that the model doesn't get overwhelmed by irrelevant features and can prioritize the most impactful connections. Together, these methods achieve the best balance between feature aggregation and interpretability. In other words, it's high-performing and provides meaningful insights into the data.

In addition, ICEWS14 likely contains latent hierarchical structures, such as political or social event patterns, which align with HierTKG's hierarchical and temporal modeling capabilities. The inclusion of TGN's memory modules further supports long-term dependency modeling, effectively capturing structured event sequences.

In comparison with our baselines, SPA (2022) and TLT-KGE (2022) rely on static or basic sequence-based embedding techniques, which struggle to capture evolving temporal and hierarchical dependencies. EmtE (2024) and TIE (2024) are advanced in handling temporal information however, these models lack HierTKG's hierarchical aggregation, limiting their effectiveness on datasets like ICEWS14 with strong latent hierarchies.

\section{Case Study}

Using the Hierarchical Temporal Knowledge Graph (HierTKG), our model can predict how a specific tweet associated with a rumor is likely to elicit responses over time. This enables the identification of chains of interactions, revealing how rumors spread from one tweet to another and through different communities on social media platforms.

For example, given a set of tweets at time $t$ about a controversial event or rumor, the model can predict which tweets are likely to be directly quoted or replied to by others at future times $t + \Delta t$. The Temporal Graph Network (TGN) captures the evolving dynamics of these tweet interactions, including the time-sensitive nature of replies or retweets. Simultaneously, the hierarchical graph pooling mechanism aggregates structural information about the larger network, such as clusters of users frequently interacting with similar content. By fusing these features, the model predicts potential future links (e.g., a user replying to or retweeting a specific tweet), enabling a detailed mapping of the rumor's trajectory across tweets.

This capability is particularly valuable for identifying tweets that may play a pivotal role in rumor amplification, such as those posted by influential users or those containing highly engaging content. By predicting which tweets are likely to spur further interactions, platforms can prioritize monitoring or fact-checking efforts on high-risk tweets. Furthermore, understanding the predicted tweet-to-tweet link patterns provides insights into how rumors transition from one conversational context to another, revealing potential cross-topic contamination or the spread of misinformation to new audiences. This targeted analysis supports proactive interventions to curb the propagation of harmful or false information.

\section{Conclusion}
In our study we introduced an algorithm HierTKG, a novel framework combining Temporal Graph Networks and DiffPooling. This approach effectively captures both the temporal evolution and hierarchical structure of knowledge graphs. It has proven to be effective, particularly for tasks like rumor propagation and temporal link prediction.

Our results highlight the impact of dataset characteristics. While structured datasets like ICEWS and WikiData align well with our model, achieving strong performance, the PHEME dataset—due to its noisy and sparse nature—presents challenges. This shows the difficulty of applying advanced models to social media datasets without additional preprocessing or augmentation.

The ablation study provided important insights. DiffPooling with Attention emerged as the best configuration, striking the ideal balance between feature aggregation and interpretability. 

The scalability and robustness of HierTKG were evident. The model’s hierarchical pooling simplifies complex graphs while preserving critical information, making it effective across a range of datasets, from fine-grained to large-scale domains.

In conclusion, our framework not only sets a strong foundation for hierarchical temporal graph modeling but also paves the way for impactful real-world applications.

\section{Future Work}
As highlighted in the Results, our model currently achieves an AP level of approximately 0.95. While other algorithms may demonstrate higher AP levels, we are confident that our novel methodology, which integrates pooling with temporal dynamics, represents a groundbreaking approach in its early stages of development. Moving forward, we aim to explore and implement advanced techniques to further enhance the AP level while simultaneously reducing the model's time complexity. Improving computational efficiency is a priority, as the current implementation requires significant runtime. These advancements will ensure the scalability and practical applicability of our model for real-world dynamic graph tasks.

\bibliography{ref}

%\bibliographystyle{plain}
%\bibliography{references}
\appendix

\section{Dataset}
% https://cs.stanford.edu/~srijan/pubs/jodie-kdd2019.pdf
% https://arxiv.org/pdf/1806.03713
% https://eventdata.parusanalytics.com/papers.dir/Arva.etal_EPSA_13.pdf

\subsection{PHEME} 
PHEME stands out for its focus on rumors and misinformation in social media, capturing how information spreads and is perceived during major events. This makes it highly relevant for research in rumor detection, misinformation, and social network analysis.

The PHEME dataset is comprised of tweets classified as rumors or non-rumors and covers nine major events, including incidents like the "Charlie Hebdo" attack, "Sydney siege," and "Ferguson" protests. It serves as the primary dataset for testing our algorithm’s ability to classify and predict rumors. Each tweet is fact-checked by professional journalists, who label rumors as either true, false, or unverified.

The following table created in \cite{kochkina-etal-2018-one} summarizes the size of each event in the PHEME data set:

\begin{table}[h]
    \centering
    \resizebox{\textwidth}{!}{%
        \begin{tabular}{|l|c|c|c|c|c|c|c|}
            \hline
            \textbf{Events} & \textbf{Threads} & \textbf{Tweets} & \textbf{Rumors} & \textbf{Non-rumors} & \textbf{True} & \textbf{False} & \textbf{Unverified} \\
            \hline
            Charlie Hebdo & 2,079 & 38,268 & 458 & 1,621 & 193 & 116 & 149 \\
            Sydney siege & 1,221 & 23,996 & 522 & 699 & 382 & 86 & 54 \\
            Ferguson & 1,143 & 24,175 & 284 & 859 & 10 & 8 & 266 \\
            Ottawa shooting & 890 & 12,284 & 470 & 420 & 329 & 72 & 69 \\
            Germanwings-crash & 469 & 4,489 & 238 & 231 & 94 & 111 & 33 \\
            Putin missing & 238 & 835 & 126 & 112 & 0 & 9 & 117 \\
            Prince Toronto & 233 & 902 & 229 & 4 & 0 & 222 & 7 \\
            Gurlitt & 138 & 179 & 61 & 77 & 59 & 0 & 2 \\
            Ebola Essien & 14 & 226 & 14 & 0 & 0 & 14 & 0 \\
            \hline
            \textbf{Total} & \textbf{6,425} & \textbf{105,354} & \textbf{2,402} & \textbf{4,023} & \textbf{1,067} & \textbf{638} & \textbf{697} \\
            \hline
        \end{tabular}%
c    }
    \caption{Number of threads, tweets, and class distribution in the PHEME dataset.}
    \label{tab:pheme_distribution}
\end{table}

%Certain events, such as "Charlie Hebdo" and "Sydney Siege", contain more tweets than other events. This difference is reflective of public interests. Additionally, accross all events, there is a significant amount of non-rumor tweets compared to rumor tweets. This suggests that despite commonly held beliefs about rumors during crises, most users shared factual information. The ratio for rumors to non-rumors vary per event and is indicative of how much information the public has on the subject. 

%Our experiments specifically focus on the "Charlie Hebdo" event due to its volume of tweets and high verification level, making it ideal for assessing rumor prediction accuracy.  This selection serves as a baseline, from which we can later generalize our findings across other events in the dataset.

\subsection{ICEWS}

The Integrated Crisis Early Warning System (ICEWS) is a comprehensive knowledge graph dataset for analyzing and predicting global conflicts \cite{lockheedmartin_icews}. It employs quantitative methods to forecast instability, with versions such as ICEWS14 and ICEWS18 covering data from 2014 and 2018, respectively. Notably, ICEWS18 has demonstrated predictive capabilities for Asian conflicts up to six months in advance \cite{wang2022search}.

\textbf{Key Features of ICEWS:}
\begin{itemize}
    \item \textbf{Event Types:} Events are categorized into several types, including:
    \begin{itemize}
        \item Domestic Crisis: Internal political tensions or social unrest.
        \item Ethnic Violence: Conflicts involving ethnic or social groups.
        \item Insurgency: Armed rebellions or violent opposition to state authority.
        \item International Crisis: Diplomatic disputes or military confrontations between nations.
        \item Rebellion: Uprisings or protests challenging political systems.
    \end{itemize}
    \item \textbf{Event Attributes:} Each event entry includes:
    \begin{itemize}
        \item \textit{Date and Location:} Specific details of the event’s occurrence.
        \item \textit{Participants:} Entities involved, such as governments or civilian groups.
        \item \textit{Event Coding:} Standardized classification of type and severity.
        \item \textit{Event Outcomes:} Details on impacts like casualties or political changes.
    \end{itemize}
\end{itemize}

ICEWS supports longitudinal analysis of conflict trends, enabling temporal modeling of event sequences. Its comprehensive dataset is invaluable for evaluating algorithms, especially for predicting non-rumor data related to global instability.

\subsection{WikiData}
WikiData, inspired by usage in the JODIE framework, is a graph dataset representing user interactions on Wikipedia. In this dataset, nodes represent Wikipedia pages, and edges represent interactions or connections between pages. WikiData allows predictions on metrics such as average daily traffic for Wikipedia pages, which is especially useful in applications focused on interaction dynamics and page popularity.

\section{Ablation Study More}

We tested the efficacy of alternative pooling mechanisms by replacing the DiffPoolLayer with either \textit{SAGPooling} or \textit{Double SAGPooling} layers.
      
        \begin{figure}[h]
            \centering
            \includegraphics[width=\linewidth]{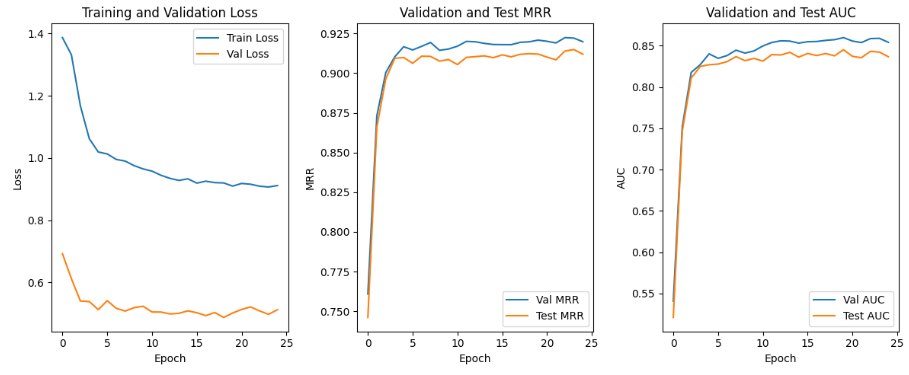}
            \caption{Ablating DiffPoolLayer by Testing with SAGPooling Only}
            \label{fig:sagpool-performance}
        \end{figure}

When SAGPooling replaced had an additional layer, the model achieved its best performance at \textbf{Epoch 33}. This result indicates that while SAGPooling is effective for feature aggregation.
        
        \begin{figure}[h]
            \centering
            \includegraphics[width=\linewidth]{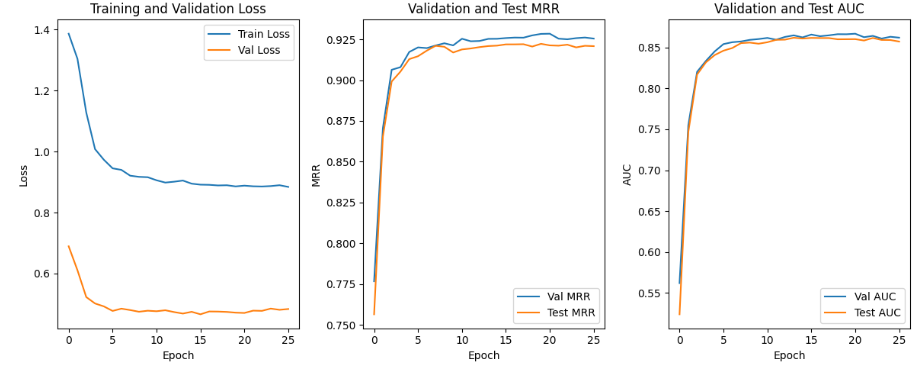}
            \caption{Ablating MemPoolLayer by Testing with Double SAGPooling}
            \label{fig:doublesagpool-performance}
        \end{figure}

We further analyzed the effect of attention-based pooling by modifying the MemPoolLayer. We compared the model's performance without multi-head attention in the \textit{FeatureFusion} module. \\
    
When SAGPooling had an additional layer without attention, the model achieved its best performance at \textbf{Epoch 38}.
        \begin{figure}[h]
            \centering
            \includegraphics[width=\linewidth]{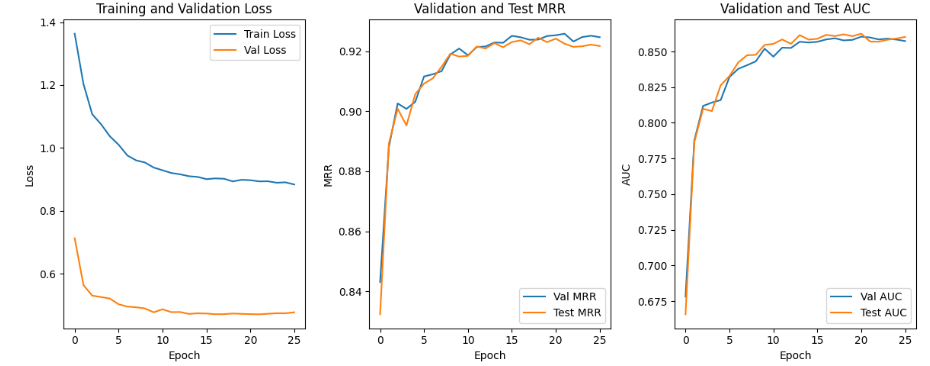}
            \caption{Ablating Attention Mechanism by Testing without Attention}
            \label{fig:sagpool-performance}
        \end{figure}

Attention-based pooling improved the validation AUC and convergence speed. Models without attention exhibited less robust feature refinement, particularly in noisy or sparsely connected graphs.

The integration of temporal memory embeddings within the pooling mechanism (MemPoolLayer) significantly outperforms traditional pooling methods such as SAGPooling and DiffPooling. This innovation effectively bridges the gap between static graph pooling and temporal representation learning. The ablation study underscores the robustness of memory-based embeddings and the critical role of multi-head attention in feature refinement. Moreover, optimizing the architecture (e.g., reducing GPU usage by employing a single-layer MemPool) enhances scalability, enabling the model to be applied to large-scale dynamic graphs.

SAGPooling consistently demonstrated better performance compared to DiffPooling, highlighting its compatibility with the given architecture. However, the MemPoolLayer's ability to provide temporal memory context proved essential for achieving optimal results. The ablation study further confirmed that removing multi-head attention negatively impacts performance, reinforcing its importance in feature fusion and the overall model's efficacy.

\end{document}